%% file: bare_jrnl.tex
\documentclass[journal]{IEEEtran}
\usepackage{xcolor}
\ifCLASSINFOpdf
\usepackage[pdftex]{graphicx}
\else
\fi

\usepackage{pgfplots}
\pgfplotsset{compat=1.15}
\usepackage{adjustbox}

\usepackage{gincltex}
\usepgfplotslibrary{fillbetween}
\usepgfplotslibrary{groupplots}
\usetikzlibrary{plotmarks}
\usetikzlibrary{patterns}
\pgfplotsset{
tick label style={font=\footnotesize},
label style={font=\footnotesize},
legend style={font=\footnotesize},
}

\definecolor{orange_D}{rgb}{1,0.3,0}%
\definecolor{cyan}{rgb}{0,0.67,0.64}%

\def \fwidth{0.95\linewidth}
\def \fheight {0.6\linewidth}

\begin{document}
%
\title{Internet of Things (IoT) Connectivity in 6G: An Interplay of Time, Space, Intelligence, and Value}

\author{Petar Popovski,~\IEEEmembership{Fellow,~IEEE}, Federico Chiariotti,~\IEEEmembership{Member,~IEEE}, Victor Croisfelt, Anders E. Kal{\o}r,~\IEEEmembership{Student Member,~IEEE}, Israel Leyva-Mayorga,~\IEEEmembership{Member,~IEEE}, Letizia Marchegiani,~\IEEEmembership{Member,~IEEE}, Shashi Raj Pandey,~\IEEEmembership{Member,~IEEE}, Beatriz Soret,~\IEEEmembership{Member,~IEEE}
\thanks{The authors are with the Department of Electronic Systems, Aalborg University Denmark. Corresponding Author: Petar Popovski, petarp@es.aau.dk.}}


\maketitle

\begin{abstract}
Internet of Things (IoT) connectivity has a prominent presence in the 5G wireless communication systems. As these systems are being deployed, there is a surge of research efforts and visions towards 6G wireless systems. In order to position the evolution of IoT within the 6G systems, this paper first takes a critical view on the way IoT connectivity is supported within 5G. Following that, the wireless IoT evolution is discussed through multiple dimensions: time, space, intelligence, and value. We also conjecture that the focus will broaden from IoT devices and their connections towards the emergence of complex IoT environments, seen as building blocks of the overall IoT ecosystem. 
\end{abstract}

\IEEEpeerreviewmaketitle

\section{Introduction}

The term \emph{Internet of Things (IoT)}, although present for several decades, started to gain a significant traction with the emergence of the 5G cellular systems and standards~\cite{palattella2016internet,8519960}. An IoT device is a physical object equipped with sensors and/or actuators, embedded computer and connectivity. As such, it can be seen as a two-way micro-tunnel between the physical and the digital world: physical information gets a digital representation and, vice versa, digitally encoded actions get materialized in the physical world. From a different perspective, related to service and product design, IoT capabilities have significantly transformed many products by expanding the functionality and transcending the traditional product boundaries~\cite{porter2014smart}.

The ambition of 5G has been to push the boundaries of connectivity beyond the offering of high wireless data rates and expand towards interconnecting humans, machines, robots, and things. This leads to an enormously complex connected ecosystem: a large number of connections that pose a vast diversity of heterogeneous Quality of Service (QoS) requirements in terms of data rate, latency, reliability, etc. To deal with this complexity, the approach of the 5G system design has been to define three generic services: eMBB (enhanced Mobile BroadBand), mMTC (massive Machine-Type Communications (MTC)), and URLLC (Ultra-Reliable Low-Latency Communication)~\cite{ITUR,tr38802}. The latter two represent the approach of 5G to natively support the requirements of IoT connectivity. This is in contrast to the 4G and other prior generations, where IoT connections were supported in an ad hoc manner, as an afterthought in system deployment.

In some sense, 5G is a step in the direction of obtaining an ultimate connectivity system that is capable to flexibly support all conceivable wireless connectivity requirements in the future. One can think of the three generic connectivity types as three dimensions of a certain ``service space'' and any single connectivity service can be realized as a suitable combination of eMBB, mMTC, and URLLC. For example, in an advanced agricultural scenario, a remotely-controlled machine needs to support real-time reliable actuation of commands (URLLC), while occasionally sending a video feed (eMBB) as well as gathering data from various sensors and IoT devices in the agricultural environment (mMTC).

\begin{figure}[t!]
\centering 
\includegraphics{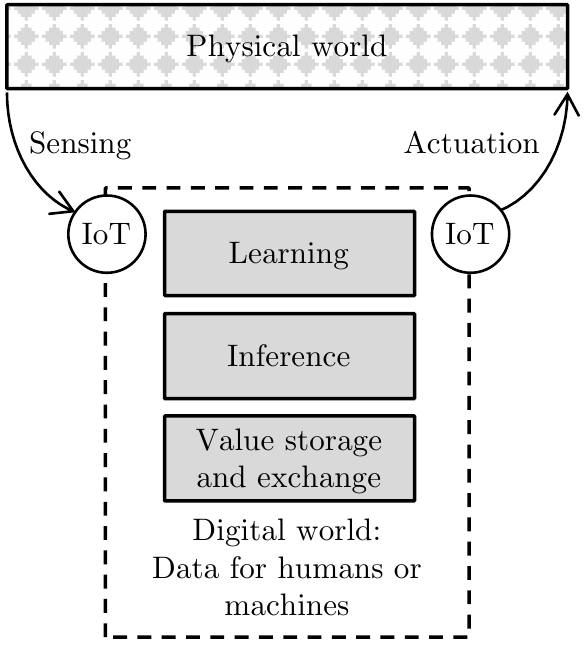}
\caption{Physical versus digital.} 
\label{fig:PhysicalVsDigital} 
\end{figure}

But is 5G indeed defining the ultimate connectivity framework? This is an important question, as its affirmative answer would obviate the need to redefine and conceptually upgrade the connectivity types towards 6G. On the contrary, a negative answer entails a critical view on 5G and identification of connectivity scenarios that are not well represented by eMBB, mMTC and URLLC or a combination thereof. 

As an attempt to answer the question above, this paper takes the, rather general, perspective depicted on Fig.~\ref{fig:PhysicalVsDigital} to position the role of IoT connectivity and assess its requirements. The general framework from Fig.~\ref{fig:PhysicalVsDigital} will be first used to take a critical view on IoT as defined in 5G and identify cases that are not well represented by the two categories mMTC and URLLC. Next, the framework will be used as a blueprint to formulate the features of IoT connectivity in beyond-5G/6G systems. The evolution of future wireless IoT technology will be discussed through multiple dimensions: time, space, intelligence, and value. We also conjecture that the focus will broaden from IoT devices and their connections towards the emergence of complex IoT environments, seen as building blocks of the overall IoT ecosystem.

\section{A General IoT Framework}

IoT devices reside at the interface between the physical and the digital world and facilitate the two types of information transfer: (1) \emph{sensing}, creating a digital representation of the physical reality and (2) \emph{actuation}, converting digital data into commands that exhibit an impact on the physical world. After the information gets converted into a digital data, it can be used in three principal ways:
\begin{itemize}
\item \emph{Learning:} The data is used in a process of training a module that relies on Machine Learning (ML) or another form of gathering knowledge and building up Artificial Intelligence (AI). 
\item \emph{Inference:} The data is used by an algorithm, AI module or similar to infer conclusions or devise a command that needs to be actuated in the physical world. 
\item \emph{Value storage or exchange:} The data is stored for potential use at a future point, such that it possesses a latent value. The data can also be exchanged through the connectivity infrastructure and thus get an actual valuation/monetization. 
\end{itemize}
The ``digital world'' box includes anything that can store, process or transfer digital data, including the the global Internet.

There are two general modes that involve IoT communication: \emph{Machine-to-Machine (M2M)}, that includes interaction and communication only among machines as well as  \emph{Machine-to-Human (H2M)} (or vice versa), where the overall IoT communication also includes a human in the loop. The principal difference between these two modes is that, when there is a human in the loop, the timing and processing constraints should conform to the ones of the human, while in the case of M2M they are subject to design and specification. In the diagram on Fig.~\ref{fig:PhysicalVsDigital}, a human belong to the physical world. In that sense, the actuation can be understood in a more general way, such as displaying a multimedia content to for the human.

\section{A Critical View on IoT in 5G}
\label{sec:Critical5G}

This section discusses the typical ways in which IoT requirements are articulated within 5G. The objective is to take a critical view by pointing out important scenarios and requirements that are not well covered by the two categories, mMTC and URLLC.

\subsection{mMTC}

\begin{figure}[t!]
\centering 
\includegraphics{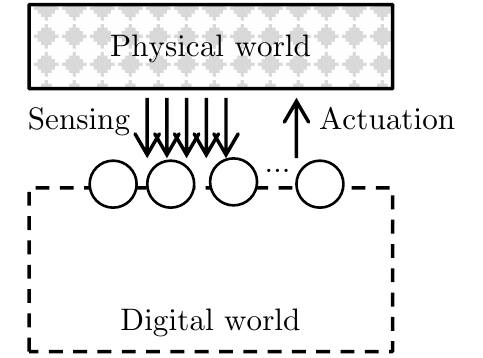}
\caption{A view on the typical mMTC requirements.} 
\label{fig:mMTC} 
\end{figure}  

We start by considering mMTC and a view on its typical requirements is depicted in Fig.~\ref{fig:mMTC}. It aims to support a large number of devices, dominantly with an uplink traffic, which is also indicated on the figure. A possible rationale behind this can be formulated as follows. Consider a large set of nodes (sensors) that are generating data locally. The data of different nodes is not correlated, such that each new data packet send by a different node contributes with a new information about the physical world. Furthermore, each node is only sporadically active, such that the time instant at which it is active and has a data to transmit is unpredictable. Equivalently, this implies that the subset of active nodes at a given time instant is unpredictable. Hence, there are two sources of randomness: the data content and the node activity. This is the basis for the major challenge in mMTC: \emph{How to maximize the uplink throughput from a large set of connected nodes, where the subset of active nodes at a given time instant is unknown?} This has led to a surge on research in the area of massive random access~\cite{polyanskiy17,bockelmann18}. The challenge requires maximization of the throughput in the uplink, which is more difficult than the downlink, as the devices are uncoordinated and compete for the same shared wireless spectrum. An additional challenge for mMTC devices is the power consumption, which should be optimized to ensure long battery lifetime and unattended operation; this is the case, for example, for sensors embedded in buildings, production plants, or agricultural facilities. More generally, the challenge for mMTC (and we will see that it is similar for URLLC) is made in a maximalist way: it is tacitly assumed that if the most difficult mode of communication is supported, then the easier modes (such as downlink communication towards a subset of nodes) are implied. Finally, following the architectural practice of layered design and modularization, the part of connectivity on Fig.~\ref{fig:mMTC} is decoupled from and oblivious towards the goals/usage of the mMTC data in the digital world, that is, learning, inference, or value storage/exchange. 

Let us now look at a scenario of massive access in which we change some of the assumptions behind the canonical mMTC use case, described above. Consider the case in which the nodes are sensing a physical phenomenon in order to sense an anomalous state and report it to an edge server, which embodies an inference module that is capable to detect reliably if an anomalous state has occurred. In the simplest case, each sensing node can make a local binary decision whether an anomalous state has occurred (1) or not (0) and send it to the edge server. This violates the assumption that the data across nodes is not correlated, as all of them will try to report about the same observed phenomenon. Furthermore, if the anomalous state occurs within a short time interval, it will trigger response from all sensing nodes in a correlated way, which impacts the statistical properties of the subset of active mMTC nodes.   In the ideal case, when a node detects the anomalous state perfectly and the wireless link to the edge server is error-free, then only a single node needs to transmit. Hence, the technical problem is not anymore \textsl{``throughput maximization from an unknown random subset''} but rather a \textsl{``leader election from an unknown subset with a certain correlation structure in the node activation''}. If the ideal assumptions on sensing/communication are relaxed, then a sufficient number of nodes should report the detected alarm, such that the edge server can infer a reliable decision about the state of the physical world. The problem can be further relaxed by considering that the sensing node are not directly detecting the anomalous state, but rather a data related to it; then the edge server needs to fuse this data to make inference. The technical problem is now \textsl{``collect a sufficient number of data points to reliably detect anomaly''}. 

These examples show that the consideration of the data purpose/usage has a significant impact on the technical challenges posed to the wireless connectivity part. For all these new technical problems, a system optimized for mMTC with typical requirements will lead either to inefficient operation (collecting much more data points than needed to make inference) or failing to meet the timing requirement (the detection of the alarm will be delayed due to channel congestion). The case of transmission of identical alarm messages from a massive set of devices that should enable timely and reliable detection at an edge node illustrates that massiveness, reliability and latency may not be separable (as treated in 5G) when the data content/purpose is taken into account. Clearly, following the (overused) approach of cross-layer optimization, one may immediately jump to the conclusion that the access should be designed jointly with the high-level objective of the transmitted data. This is not feasible, as it does not contribute to a scalable architectural design. However, the described problems indicate that, rather than sticking to the problem of \textsl{``throughput maximization from an unknown random subset''}, we need to identify a small set of connectivity-related challenges that provide a better span of the IoT requirements with massive number of devices and design systems that can solve them efficiently. 

\subsection{URLLC}

\begin{figure}[t!]
\centering 
\includegraphics{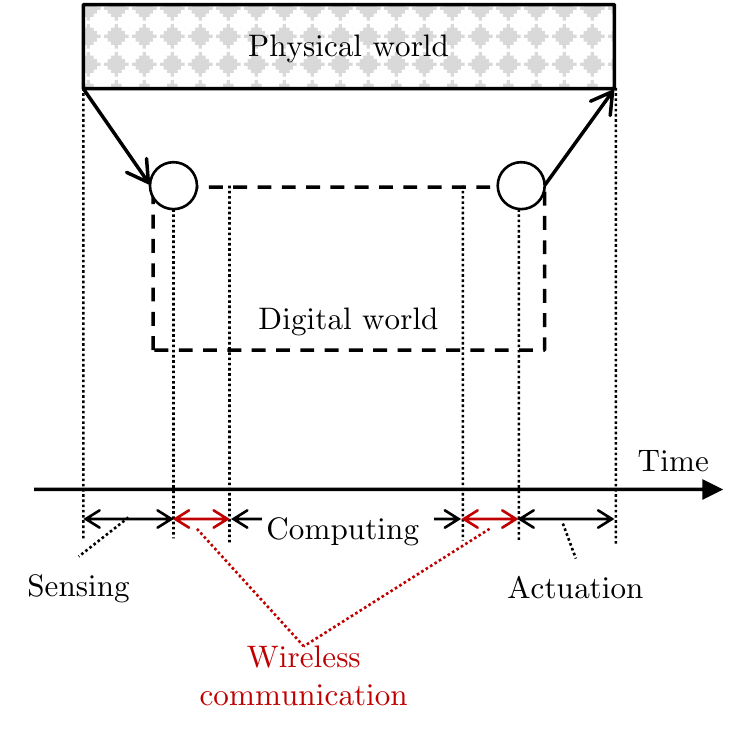}
\caption{The context for defining of URLLC requirements through a timing budget of a sense-compute-actuate cycle.}
\label{fig:URLLC} 
\end{figure}

We now provide a critical view on URLLC, the second generic service related to IoT. In order to illustrate the URLLC requirements, we consider the sense-compute-actuate cycle depicted on Fig.~\ref{fig:URLLC}. In this example we observe the timing of the following loop. An IoT gathers information from the physical world, digitalizes it and transmits it wirelessly to a server that performs computation and inference. Based on that, the server sends a command wirelessly to an actuating device; in the special case, this device is the same one as the sensing IoT device. Fig.~\ref{fig:URLLC} illustrates the total timing budget for these operations. The specification of URLLC has been done with the motivation to use a small part of this timing budget on the wireless radio link and ensure that transmission is done within this short allocated time with a very high reliability. This would leave a sufficient timing budget to perform the other operations, regardless of whether the total timing budget is $10$ ms or $50$ ms.  

This is again a rather maximalist approach towards the radio link in the quest to satisfy the end-to-end requirements on timing and reliability. In the early specification of URLLC, the allocated time was $1$ ms and the required reliability was $99.999$\%. Achieving high reliability is associated with the use of high level of diversity (e.g. bandwidth) and power. Relaxing the requirements on the wireless transmission could lead to more efficient operation, while still meeting the overall goal of communication. Specifically, in the example on Fig.~\ref{fig:URLLC} the computation part may be capable to compensate for the data loss on the sensing wireless link and make a predictive decision that can be passed on to the actuator. Or, as indicated in the early paper on ultra-reliable communication on 5G~\cite{Petar2014}, one can take a holistic perspective on URLLC and ensure that the overall system degrades gracefully if the data is not delivered within a given deadline. 

Expressing it in a similar way as we have done in the previous section, the basic problem of URLLC has been formulated as \textsl{``deliver the data of size $X$ within $Y$ milliseconds with reliability $Z$''}. Instead, timing in a communication system can be put in a more general framework and define a set of basic problems that are capable to capture various timing requirements. For instance, instead of looking at the latency of the packet, one can jointly consider the data generation process and the state of the computation process. In that sense, a more relevant measure than latency can be information freshness or age of information. This is further discussed in Section~\ref{sec:Time}, while for a more general discussion on the timing concepts towards 6G the reader is referred to~\cite{popovski21timing}.

\section{Time}\label{sec:Time}

The definition of \emph{real time} is highly dependent on the application and its final user. Specifically, real time is dependent on whether the overall system is intended for one of the following three communication setups: (1) Human-to-Human (H2H), (2) Human-to-Machine (H2M), including setups of communication among machines with a Human In The Loop (HITL); and (3) Machine-to-Machine (M2M). Even fully interactive human-type communication such as Augmented and Virtual Reality (AR/VR) do not require millisecond timescales, as human perception becomes the limiting factor~\cite{miller1968response}: For example, the human eye cannot perceive images that are shown for less than about 13~ms, setting a hard ceiling on the network timing requirements for this type of communication. The same is true in  HITL scenarios, where machines can operate faster than human perception limits, but the a system operating faster than the perceivable latency threshold will be experienced by the human as instantaneous and seamless~\cite{kasgari2019human}. 

There is no universally defined timing threshold for M2M communication, as the timing requirements depend on the type of applications and on the capabilities of the specific Cyber-Physical System (CPS). This is in a way reflected in the split between mMTC and URLLC in 5G, which represent two extreme cases. As also discussed in Section~\ref{sec:Critical5G}, these two extremes do not cover the full range of use cases. A different and representative categorization in terms of timing is given by the OpenRAN Alliance (ORAN)~\cite{oranrt}, which distinguishes three main classes of traffic: non real-time, near real-time, and real-time. A more accurate view of CPS timing requirements should go beyond the isolated characterization of the wireless communication latency and consider all the contributors in Fig.~\ref{fig:URLLC}. Furthermore, the use of Age of Information (AoI)~\cite{kaul2012real} instead of latency as a metric can have significant advantages, as AoI can better represent the discrepancy between the model that the system can construct from the sensor transmissions and the actual physical reality. The limits on the allowed AoI depend on the tolerance of the control algorithm and of the application: advanced control algorithms in highly predictable scenarios will be able to work even with very old information, while complex and unpredictable scenarios which require fast reaction times will necessarily have stricter requirements~\cite{zhang2019networked}. 

\begin{figure}[t]
 	\centering
    \input{fig/aoi.tex}
    \input{fig/voi.tex}
  \caption{Example of the difference between AoI and VoI, in a system with cumulative estimation errors.}
  \label{fig:aoi_vs_voi}
\end{figure}
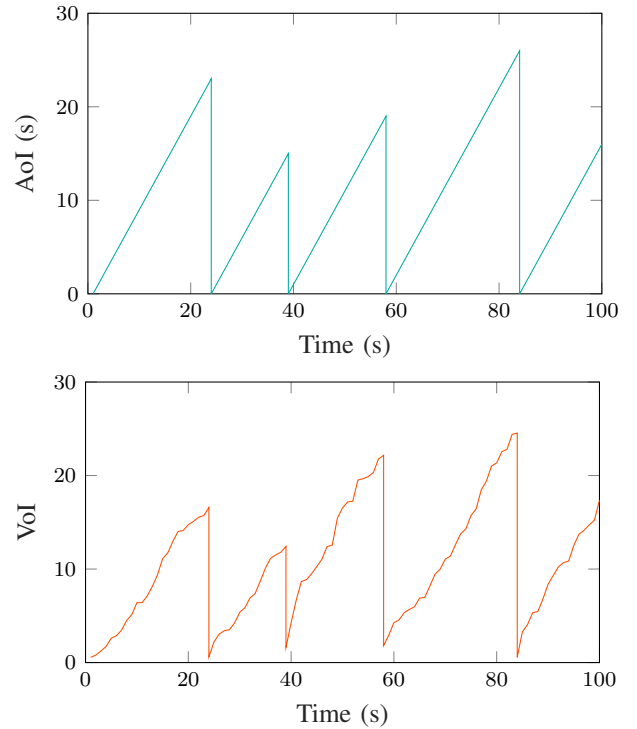

One step further is to use the content of the packets themselves to define latency and reliability requirements: if the controller employs some form of predictive algorithm, new information that fits the expected model will be relatively unimportant, while unexpected deviations from it will need to be delivered quickly and reliably. This approach can be measured with the Value of Information (VoI)~\cite{ayan2019age}, a metric that combines the age and content of the packet to directly measure the usefulness of communication. The difference between AoI and VoI is clearly shown in Fig.~\ref{fig:aoi_vs_voi}: While AoI increases linearly and then drops to 0 after a transmission (assuming the latency is negligible), the increase in the VoI depends on the behavior of the system, and might be non-linear. In the figure, the first period between 0 and 25~s has a relatively slow increase, while the period between 40 and 60~s has a steeper one, and indeed gets to a higher VoI in a shorter time: this is due to the different behavior of the system, which strays farther from the estimated value at the receiver.

Using VoI as a metric is a step towards \emph{semantics}-oriented communication. The classical design of a CPS assumes a total independence between the content of the data and their transmission, i.e.,  uncontrolled arrival of exogenous traffic to the communication system. This sets design boundaries to the communication protocols and relaxing this rigid separation allows us to tackle the system design process holistically and improve the performance.
In control and HITL applications the Urgency of Information (UoI) approach~\cite{zheng2020urgency} defines VoI in such a way that the packets with the highest value are the ones that affect the control performance the most. This definition of value is also closely tied to the market value of data, which we will describe below: in both cases, samples from the sensors are more valuable if they are \emph{surprising}, i.e., if they contain information that is not currently represented in the model of the system. The difference between the two is in the way the data is used: in the data market case, this new information is used to improve that model, while in VoI applications, it is used to track and control a system.

\section{Space}

Evolving towards 6G, we need to look in the changes that occur in the space in which IoT devices are deployed to operate. In this context, we use the term space to refer to: (1) the environment where sensing and actuation take place and (2) the propagation medium where the electromagnetic waves travel to transfer information between two or more points. 
Hence, by delimiting the space in which the interface between the physical and digital worlds  occurs, the definition of time (space-time) and frequency as resources for communication is inherent. Therefore, space sets the basis for resource sharing and competition among devices.

As the optimization of frequency and time resources becomes insufficient, the next frontier towards increased network capacity is to optimize the use of space. Thence, increasing the network capacity per unit area has been one the major objectives of every subsequent generation of mobile networks. However, this objective has encountered a major challenge: the optimization has been limited to the placement and capabilities of the networking devices -- the infrastructure -- whereas there has been little to no control on the user side. That is, IoT and other mobile user devices possess limited capabilities and, as a consequence, their wireless channel is mostly determined by nature. Because of this, the traditional approach towards a greater network capacity is pre-planned network densification in combination with frequency reuse to minimize inter-cell interference. Only in recent years, precoding, beamforming, and beam steering techniques have enabled a much more flexible and agile exploitation of the space resources through massive multiple input-multiple-output (mMIMO) \cite{Marzetta2010} and the development of cell-free networks \cite{Demir2021}. In mMIMO, the channel state information, based on spatial reciprocity, is exploited to achieve communication with multiple devices in the same block of time-frequency resources.
Furthermore, distributed or cell-free mMIMO allows to exploit the macro-diversity of the environment by allowing the IoT devices to communicate to antenna elements at different locations to combat blockages and eliminate coverage holes.

Despite these advances, the capabilities of the IoT devices will remain limited in order to keep their cost down. Nevertheless, new developments on distributed infrastructures, AI/ML, and signal processing techniques will enable the network infrastructure and the environment itself to become intelligent. This will enable the real-time  self-optimization of heterogeneous architectures that can relax the hardware requirements of IoT devices while exploiting the spatial resources. In the recent developments, the propagation environment may act as an ally to the simple IoT devices rather than only a major challenge that needs to be overcome. For instance, reconfigurable intelligent surfaces (RIS) \cite{Bjornson2021} consist of elements that can alter the properties of the incident signals adaptively and, hence, allow for a much greater control over space than that of the IoT devices. Specifically, RIS can be used to take advantage of the location of the devices to create highly directive and interference-free beams towards the base station in real-time. This allows for a new interpretation and exploitation of overlapping signals and also alleviates the hardware requirements of the devices, since part of the hardware on the device can be outsourced to the environment. 

The structure of the physical space plays a major role on the feasibility of deploying network infrastructure and, hence, on the availability of Internet connectivity. Historically, we have seen the infrastructure as deployed in the 2D space; usually encompassing ground-level infrastructure while considering the air and (outer) space infrastructure to be, oftentimes, alien to it and, in the best case, complementary (i.e., global positioning and navigation services). It is only in recent years, that the New Space era and the advances in unmanned aerial vehicles (UAV) have expanded our view of the network infrastructure to the 3D space~\cite{Kodheli2021}. Satellites deployed in the Low Earth Orbit (LEO) can serve as a global network, capable of achieving low end-to-end latency while providing coverage in remote regions (e.g., Arctic and maritime) where deploying terrestrial infrastructure is infeasible~\cite{abildgaard2021arctic}.  Besides, while LEO satellite constellations are moving infrastructure, they present deterministic space-time dynamics that can be exploited for resource optimization~\cite{Leyva-Mayorga2021}. Due to this combination of characteristics, one of the major objectives for 6G is to achieve a full integration of the terrestrial infrastructure with satellites, drones, and other aerial devices to fully exploit the 3D nature of space~\cite{Di2019, Dang2020, Akyildiz2020}.

In the digital world, space influences a series of characteristics of the data beyond quantity, such as its content and, hence, relevance. This calls for a characterization of how the optimization of wireless resources for a given delimited space affects the overall learning, inference, and value of data. In this sense, the network-level optimization objectives must be redefined to consider the role that space plays in the use that the data will have.  
Consequently, there is the need for a new interpretation of resource efficiency 
depending on the context: \emph{What is the data content and what will it be used for?} For example, the concept of over-the-air computation exploits the superposition property of the medium to effectively merge data from multiple sensors or model updates in the case of distributed learning \cite{Park2021}. This indicates that the 5G interpretation of space is far from being a definitive vision, as the capitalization of space, now dynamically, depends mainly on the utility of the data.

\section{Intelligence: Learning and Inference}

A general and rather certain trend in the coming years is that the intelligence in networks, network nodes, but also connected devices, will continuously increase. As the number of applications relying on IoT technology has grown, we have seen the capabilities of those \textit{things} evolving accordingly. Devices that used to be exclusively employed as sense-and-transmit entities are now equipped with different levels of \textit{embedded intelligence} directly operating on the information collected. This need for increasingly smarter communicating parties opens the way to the definition of a ``smarter'' content to exchange and of novel ways of making sure this is done efficiently and correctly. There have been already some efforts in this perspective \cite{zalewski2020bits, elsts2018board}, where the communication effort is optimised so that only the most useful data for the actual data consumer is transmitted. In a machine learning  perspective this could, for instance, get translated into ``communicate only the most significant features''. But what is actually determining the significance of the information exchanged in this context? And how do we make sure this is transmitted efficiently? One natural option would be to consider relevant whatever maximises the performance of the receiver at executing a specific task, while relying an a compression strategy able to extract exactly this relevant information from the data. Communication frameworks based on neural network autoencoders to encode and decode messages would fulfil the requirements described above, as inherently able to find compressed input representations which are the most useful for the task at hand (e.g., \cite{bourtsoulatze19}). Yet, this is not enough. Systems would end up being \textit{ad-hoc} systems, able to operate correctly only on a small set of tasks (by leveraging multi-task learning schemes \cite{vandenhende2020revisiting}; otherwise only on a single task), where all the parties involved in the communication share the same model (i.e., same network with same structure, parameters and weight values), and interpretability would remain a crucial challenge. Preliminary studies have shown the potential of data-driven techniques (such as autoencoders) in this context \cite{xie21, chen2019toward}; though, it is evident how such paradigms create strong constraints against generalisation. This is why we will eventually need to move away from those systems and start associating \textit{semantics and meaning} to the data, as also suggested by a series of recent papers on semantic communication~\cite{popovski2020semantic,kountouris21semantics,uysal21semantic} as well as advances in graph and semantic networks \cite{wang2019heterogeneous, zhao2019semantic}. Indeed, by integrating semantics-based systems (i.e., systems with knowledge representations, such as knowledge graphs, and reasoning capabilities), well studied and long exploited in more traditional Artificial Intelligence (AI), with more recent data-driven frameworks,  we would eventually enable efficient communication of relevant and meaningful information among entities sharing the same view of the world in the form of a \textit{knowledge base}, rather than a specific single, task-dependent, model~\cite{lan21semantic}.

\section{Value}

The inter-networked CPSs in the IoT networks are readily accumulating and processing data at a large scale. When operated with Machine Learning (ML) tools, these massively distributed data stimulate \textit{real-time} and \textit{non-real time} inference and decision-making services that create an economic value of data. For example, services defining prediction, localization, automation and control heavily consume large data samples for training learning models and improve its performance, i.e., model accuracy. These are certainly a few of the several promising outlooks with data in general. In particular, data is a valued commodity for trade that has gathered significant economic value and a multitude of social impacts. However, it is an overstatement if the narratives on the economic value of data leave the fundamental inter-dependencies between the data properties itself, the contextual, time-space information it encodes, and its value.

\begin{figure}[t!]
\centering 
\includegraphics[width=\linewidth]{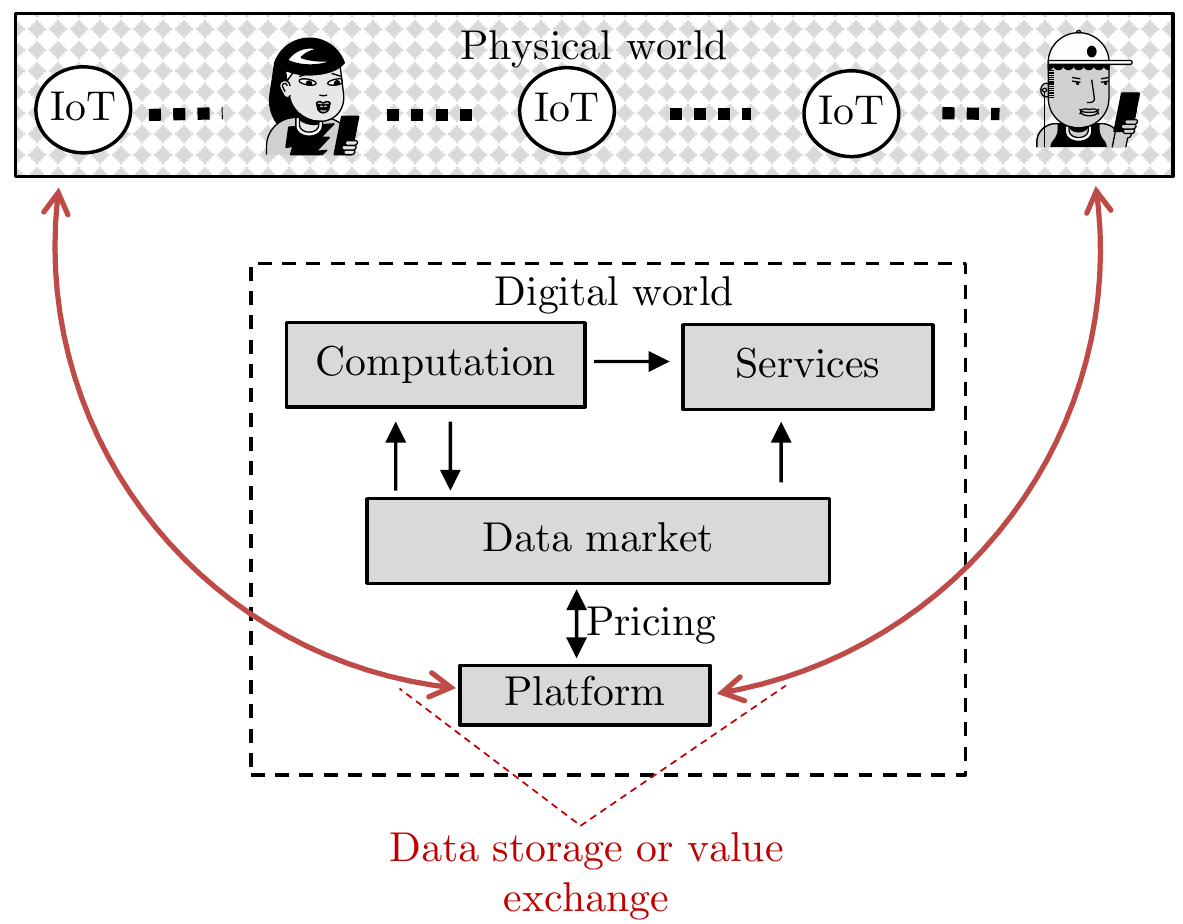}
\caption{An illustration of data trading in an IoT network.}
\label{fig:value} 
\end{figure}

At the other end of the story, the \textit{utility} of distributed data in the physical space is constantly challenged by a conventional perspective at an IoT device. The hindrance resides in the fundamental attributes of IoT networks and their components: data are not readily usable and highly localized, the data sources are resource-constrained, and the system is under stress due to unreliable connectivity during data transfer. Whereas, data in the digital space permits flexibility in its storage, mobility, customization and inter-operation to extract meaningful information, behaving likewise digital goods. Therein, data can be monetized and exchanged for added value, as shown in Fig.~\ref{fig:value}. The platform is a primary interface of interaction between buyers/sellers, leveraging connectivity for value storage or exchange in the data market, which quantifies pricing schemes and facilitates the overall data trading process. In this matter, one must not confuse ``value" and ``pricing"; for instance, the utility of correlated IoT data diminishes quickly if it exhibits no latent value~\cite{ali2020voluntary, agarwal2020towards}. Hence, the monetary value of such data appears low. However, such data can still contribute to assess system-level reliability, such as in sensory networks, where IoT devices constantly transmit their measurement data, or in case the data is used for inference. Such use cases highlight the challenges of a holistic approach in quantifying the data value. However, the value storage and exchange should not be a naive characterization by the single arbitrator/platform but depends entirely on the nature (independent or collective) and the requirements of applications these data can offer. For instance, a more tailored mobile application that benefits users' with specific personalized services expects techniques to handle data privacy concerns, for which the value exchange mechanism would be unique.

Arguably, this departure in the understanding of IoT devices, basically confined within sensing/actuation functionalities, to a broader physical and digital world perspective, in principle, impacts how connectivity shall behave and value is added with data exchanges. One can think of emergence of IoT devices that will behave as autonomous sellers and buyers of data in a decentralized data market. An example, the elastic computational operation on data in the digital space, coupled with value storage or exchange technologies, such as distributed ledger technologies (DLTs)~\cite{fernandez2018review}, quantifies the utility of data as transaction details and provides a different take on communication requirements to operate data trading. Similarly, in a Smart Factory setting, the value of exchanged data between devices during operation also reveals the properties of shared media access patterns, which can be exploited as feedback to tune vital parameters defining the communication resources in general. This explains the rationale behind the need to incorporate frequent interactions between the physical and the digital world, which bring value out of data, its storage and exchange while optimizing connectivity. 

\section{Towards complex IoT environments}
Although the initial IoT designs focused on simple applications, the maturity of the technology leads towards more complex systems where the single device model falls short. Rather than isolated and low-capacity devices, we encounter IoT applications that are deployed and executed in several heterogeneous edge devices, interconnected with a network -- wireless and/or wired -- that dynamically adapts to changes in the environment and with built-in intelligence and trustworthiness. These environments rely on several distributed technologies, such as edge computing~\cite{Alnoman2019edge}, edge intelligence~\cite{Deng2020edgeintelligence} and DLT~\cite{fernandez2018review}, and their complex interactions. 

For instance, in a manufacturing plant we find a number of inter-connected industrial robot arms, machinery and Automated Guided Vehicles (AGVs). The accomplishment of a complex manufacturing goal (the IoT application) is based on the autonomous collaboration between the nodes, with very heterogeneous capabilities. This requires the orchestration of the computation and communication resources for an overall reliable, trustworthy and safe operation.  
Another example is a fleet of e-tractors equipped with sensors and computing resources to perform the mission assigned to them (e.g., harvest field ``X"). The computing resources enable each tractor to perform computation tasks on spot, thus acting as an edge-based device, and they collaborate to achieve the common goal. Tasks that cannot be performed on the vehicle will be offloaded to an available cloud infrastructure. Components within the tractor are usually connected with Time Sensitive Networking (TSN), whereas the edge-cloud communication and the tractor-to-tractor communication is wireless.

Characterizing the performance and the energy efficiency of these complex systems is a daunting task. The conventional approach has been to characterize every single device or link and technology, but this approach is too simplistic. For example, the energy expenditure of an IoT device will strongly depend on the context in which it is put, in terms of, e.g., goal of the communication or traffic behaviour. Therefore, the system performance and the total energy footprint is not just a simple sum of an average per-link or per-transaction contribution of an isolated device. A more accurate picture of the overall performance and energy consumption is obtained by taking the complex IoT system as the basic building block. At the same time, the timing characterization of the system becomes more involved, and the new ecosystem of timing metrics discussed in Section~\ref{sec:Time} must be adapted to capture the distributed interrelations~\cite{popovski21timing}.  

\section{Conclusion}

This paper has provided a perspective on the evolution of wireless IoT connectivity in 6G wireless systems. In order to justify the enthusiasm towards developing new 6G systems, we have taken a critical view on 5G IoT connectivity. Specifically, we have illustrated cases that are potentially not captured by the 5G classification of IoT into mMTC and URLLC, respectively. In order to put the IoT evolution in a proper perspective, we have started from a general IoT framework in which we identify three principal uses of the data transferred from/to IoT devices: learning, inference, and data storage/exchange. This general framework has been expanded through different dimensions of wireless IoT evolution: 
time, space, intelligence, and value. Finally, we have discussed the emergence of complex IoT environments, seen as building blocks that are suitable to analyze the energy efficiency of these systems.

\input{bare_jrnl.bbl}

\end{document}

%% file: fig/aoi.tex
%
%
%
\begin{tikzpicture}

\begin{axis}[%
width=\fwidth,
height=\fheight,
xmin=0,
xmax=100,
xlabel style={font=\color{white!15!black}},
xlabel={Time (s)},
ymin=0,
ymax=30,
ylabel style={font=\color{white!15!black}},
ylabel={AoI (s)},
axis background/.style={fill=white}
]
\addplot [color=cyan, forget plot]
  table[row sep=crcr]{%
1	0\\
2	1\\
3	2\\
4	3\\
5	4\\
6	5\\
7	6\\
8	7\\
9	8\\
10	9\\
11	10\\
12	11\\
13	12\\
14	13\\
15	14\\
16	15\\
17	16\\
18	17\\
19	18\\
20	19\\
21	20\\
22	21\\
23	22\\
24	23\\
24	0\\
25	1\\
26	2\\
27	3\\
28	4\\
29	5\\
30	6\\
31	7\\
32	8\\
33	9\\
34	10\\
35	11\\
36	12\\
37	13\\
38	14\\
39	15\\
39	0\\
40	1\\
41	2\\
42	3\\
43	4\\
44	5\\
45	6\\
46	7\\
47	8\\
48	9\\
49	10\\
50	11\\
51	12\\
52	13\\
53	14\\
54	15\\
55	16\\
56	17\\
57	18\\
58	19\\
58	0\\
59	1\\
60	2\\
61	3\\
62	4\\
63	5\\
64	6\\
65	7\\
66	8\\
67	9\\
68	10\\
69	11\\
70	12\\
71	13\\
72	14\\
73	15\\
74	16\\
75	17\\
76	18\\
77	19\\
78	20\\
79	21\\
80	22\\
81	23\\
82	24\\
83	25\\
84	26\\
84	0\\
85	1\\
86	2\\
87	3\\
88	4\\
89	5\\
90	6\\
91	7\\
92	8\\
93	9\\
94	10\\
95	11\\
96	12\\
97	13\\
98	14\\
99	15\\
100	16\\
};
\end{axis}
\end{tikzpicture}%

%% file: fig/voi.tex
%
%
\begin{tikzpicture}

\begin{axis}[%
width=\fwidth,
height=\fheight,
xmin=0,
xmax=100,
xlabel style={font=\color{white!15!black}},
xlabel={Time (s)},
ymin=0,
ymax=30,
ylabel style={font=\color{white!15!black}},
ylabel={VoI},
axis background/.style={fill=white}
]
\addplot [color=orange_D, forget plot]
  table[row sep=crcr]{%
1	0.561098976223516\\
2	0.798611512161924\\
3	1.23734947268597\\
4	1.72284798226285\\
5	2.61121119171698\\
6	2.87217224139737\\
7	3.46314139269515\\
8	4.50255141472715\\
9	5.1211510237532\\
10	6.40273319517871\\
11	6.40928899287179\\
12	7.11930703461195\\
13	8.14562947672469\\
14	9.37830475129107\\
15	11.1172904359432\\
16	11.7284417723115\\
17	12.9681317273288\\
18	14.0126596385962\\
19	14.1191737282896\\
20	14.7463273250471\\
21	15.0921926915844\\
22	15.5438954583125\\
23	15.7182686597345\\
24	16.5878141301932\\
24	0.620148005811998\\
25	2.20781299735102\\
26	3.0434433419392\\
27	3.40392701213529\\
28	3.51683640251523\\
29	4.23772424711146\\
30	5.37118230849197\\
31	5.86099587114935\\
32	6.90393138425416\\
33	7.39115376407472\\
34	8.70900193022114\\
35	10.0941500693763\\
36	11.152146455885\\
37	11.5043698665155\\
38	11.8024638911708\\
39	12.4197567801933\\
39	1.61598700196854\\
40	4.24456164764831\\
41	6.689853916808\\
42	8.66926686630751\\
43	8.87257223821188\\
44	9.5168981959421\\
45	10.272499682921\\
46	11.0553323755257\\
47	12.4084511957616\\
48	12.5537215677127\\
49	15.440733946126\\
50	16.5360804586446\\
51	17.1734599737956\\
52	17.2576349953697\\
53	19.5186908700709\\
54	19.6630905625835\\
55	19.8812270590243\\
56	20.3130734959457\\
57	21.772058811974\\
58	22.1651899705694\\
58	1.81375038245922\\
59	2.92985273931506\\
60	4.26353682014342\\
61	4.56105608837663\\
62	5.32881753580063\\
63	5.68617377062226\\
64	5.96759058398794\\
65	6.89017572045425\\
66	6.96853257546468\\
67	8.17981693656605\\
68	9.44159027850525\\
69	9.98092535096855\\
70	11.0644096460868\\
71	11.3910331878541\\
72	12.6339527796385\\
73	13.7787800855023\\
74	14.2882312537022\\
75	15.7317121991519\\
76	16.4240758481349\\
77	18.4673028957129\\
78	19.4051973310192\\
79	21.0114884796053\\
80	21.3549166599156\\
81	22.5508437799896\\
82	22.7983558478316\\
83	24.4006687728183\\
84	24.5411019912666\\
84	0.565262427417168\\
85	3.28082284843576\\
86	4.05738729352085\\
87	5.33658698876002\\
88	5.45146682161937\\
89	6.79046098360511\\
90	8.38845341026391\\
91	9.29487367324498\\
92	10.2369098431777\\
93	10.6872613925116\\
94	10.8439742355091\\
95	12.5124605370695\\
96	13.7249978108339\\
97	14.1466930730918\\
98	14.7201567506261\\
99	15.2081282037247\\
100	17.365162430292\\
};
\end{axis}

\end{tikzpicture}%